\documentclass[a4paper,11pt]{article}
\usepackage{latexsym}
\usepackage{amssymb}
\usepackage{amsmath}
\usepackage{cite}
\usepackage{color}

\usepackage{microtype}

\numberwithin{equation}{section}
\allowdisplaybreaks

\makeatletter \@addtoreset{equation}{section} \makeatother

\usepackage{ytableau}

\ytableausetup{centertableaux, mathmode, smalltableaux}

\definecolor{blue-violet}{rgb}{0.54, 0.17, 0.89}
\definecolor{PineGreen}{cmyk}{0.92, 0, 0.59, 0.25}
\definecolor{YellowOrange}{cmyk}{0, 0.42, 1, 0}

\newcommand{\be}{\begin{equation}}
\newcommand{\ee}{\end{equation}}
\newcommand{\beq} {\begin{equation}}
\newcommand{\eeq} {\end{equation}}
\newcommand{\ba}{\begin{eqnarray}}
\newcommand{\ea}{\end{eqnarray}}

\usepackage{graphicx}

\usepackage{hyperref}

\begin{document}
\numberwithin{equation}{section}

\begin{center}
{\bf\LARGE On the dilation current in metric-affine gravity} \\
\vskip 2 cm
{\bf \large D. Kenzhalin$^{1,2}$, S. Myrzakul$^{1,2}$, R. Myrzakulov$^{1,2}$, L. Ravera$^{3,4,5}$}
\vskip 8mm
 \end{center}
\noindent {\small $^{1}$ \it LN Gumilyov Eurasian National University, Nur-Sultan, 010008, Kazakhstan. \\
$^{2}$ \it Ratbay Myrzakulov Eurasian International Centre for Theoretical Physics, Nur-Sultan, 010009, Kazakhstan. \\
$^{3}$ \it DISAT, Politecnico di Torino -- PoliTo, Corso Duca degli Abruzzi 24, 10129 Torino, Italy. \\
$^{4}$ \it Istituto Nazionale di Fisica Nucleare, Section of Torino -- INFN, Via P. Giuria 1, 10125 Torino, Italy. \\
$^{5}$ \it Grupo de Investigación en Física Teórica -- GIFT, Concepción, Chile.
}

\begin{center}
\today
\end{center}

\vskip 2 cm
\begin{center}
{\small {\bf Abstract}}
\end{center}

We review $F(R,\mathcal{D})$ gravity in the metric-affine framework, where $\mathcal{D}$ is the divergence of the dilation current appearing in the hypermomentum tensor. We assume only linear couplings between the general affine connection and the matter fields (minimal coupling) and break projective invariance to preserve a nonvanishing dilation current. For $F(R,\mathcal{D})$ linear in $\mathcal{D}$ the dilation current dependence in the function $F(R,\mathcal{D})$ does not contribute to the field equations of the theory. 

We show that, on the other hand, in more complicated cases (e.g., considering the function $F(R,\mathcal{D})=R+\alpha \mathcal{D}^2$), the $\mathcal{D}$ contribution to the metric field equations is nontrivial and can affect the cosmology of the theory.

\vfill
\noindent {\small{\it
    E-mail:  \\
{\tt dkenzhalin@gmail.com}; \\
{\tt srmyrzakul@gmail.com}; \\
{\tt rmyrzakulov@gmail.com}; \\
{\tt lucrezia.ravera@polito.it.}}}
   \eject

	


\section{Introduction}\label{intro}

A wide part of the physics scientific community claims that considering generalizations and extension of Riemannian geometry may lead to a clearer understanding of open issues that still afflict general relativity. In this context, metric-affine gravity (MAG) theories \cite{Hehl:1994ue,Vitagliano:2010sr,Iosifidis:2019dua} play a prominent role, as they are based on non-Riemannian geometry \cite{Eisenhart,Klemm:2018bil}, where one allows for non-vanishing torsion and non-metricity, along with curvature. Even if MAG has been originally introduced to interpret gravity as a ``gauge'' theory, there is no conceptual or physical problem in studying metric-affine theories outside this realm (see, e.g., \cite{Iosifidis:2021iuw,Iosifidis:2021fnq,Gialamas:2022xtt,Gialamas:2023aim,Gialamas:2023emn} for recent applications). 
In particular, MAGs in the first order formalism are gravitational theories alternative to general relativity where the metric and the general affine connection (i.e., involving, in principle, torsion and non-metricity)
are considered, a priori, as independent.

Modified theories of gravity \cite{Saridakis:2021lqd}, such as the well-known family of $F(R)$ gravity theories \cite{Sotiriou:2006hs,Sotiriou:2008rp} (where each model is defined by a different function, $F$, of the curvature scalar $R$), and its extensions to more complicated cases involving different scalars, have been analyzed in the MAG framework (see, e.g., \cite{Iosifidis:2019dua,Capozziello:2009mq,Iosifidis:2020dck,Iosifidis:2021xdx}).
In the MAG setup, coupling with matter is particularly interesting to be studied, as the matter Lagrangian is considered to depend on the connection as well: When varying the matter part of the action with respect to the connection, the so-called hypermomentum tensor \cite{Hehl:1976kt,Hehl:1976kt2,Hehl:1976kv,Obukhov:1996pf,Iosifidis:2020upr} comes into play, encompassing the microscopic characteristics of matter. The energy-momentum tensor sources spacetime curvature through the metric field equations, while the
hypermomentum is source of spacetime torsion and non-metricity by means of the connection field
equations. 

A particular extension of $F(R)$ gravity is $F(R,\mathcal{T})$ gravity \cite{Harko:2011kv,BarrientosO:2014mys,Harko:2014gwa}, where $\mathcal{T}$ is the trace of the (metrical) energy-momentum tensor $T_{\mu \nu}$, which has been rather recently studied in the Palatini formulation \cite{Wu:2018idg} and in the metric-affine framework \cite{Barrientos:2018cnx}.\footnote{The literature on $F(R)$ and $F(R,\mathcal{T})$ gravity is vast. Here we report the original works and some concise reviews with further motivations and applications.} Physically, the dependence on $\mathcal{T}$ may be induced by exotic imperfect fluids or quantum effects (conformal anomaly).
An interesting cosmological motivation for $F(R,\mathcal{T})$ gravity is that it may be considered a relativistically covariant
model of interacting dark energy \cite{Harko:2011kv}. Moreover, it is thought that $F(R,\mathcal{T})$ gravity may give some hints for the existence of an effective classical description of the quantum properties of gravity (see \cite{Wu:2018idg} and references therein).

In the metric-affine framework, as we will review in the following, the divergence of the dilation current appearing in the hypermomentum tensor is what relates the trace of the so-called canonical and metrical energy-momentum tensors. Hence we can say that the energy-momentum trace $\mathcal{T}$ and the divergence of the dilation current are on the same level, in the sense that the former is the analog of $\mathcal{T}$, but for the hypermomentum tensor. As a result, a particular class of MAG theories that deserves further attention is $F(R,\mathcal{D})$ gravity, the ``hypermomentum analog'' of $F(R,\mathcal{T})$ gravity, where $\mathcal{D}$ is the divergence of the dilation current appearing in the hypermomentum tensor.

Thus, in this work we focus on metric-affine $F(R,\mathcal{D})$ gravity, which was introduced in \cite{Iosifidis:2021kqo,Myrzakulov:2021vel}. For $F(R,\mathcal{D})$ linear in $\mathcal{D}$, the dilation current dependence in the function $F(R,\mathcal{D})$ does not contribute to the field equations, as $\sqrt{-g} \mathcal{D}$ in the action is a total derivative. However, in more complicated cases (e.g., considering a function $F(R,\mathcal{D})$ that is linear in $R$ but quadratic in $\mathcal{D}$: $F(R,\mathcal{D})=R+\alpha \mathcal{D}^2$), the contribution from $\mathcal{D}$ in the metric field equations is nontrivial and also affects the cosmology of the theory. In particular, as we will see, the presence of the dilation current $\mathcal{D}$ affects the first (modified) Friedmann equation.

The remainder of this paper is structured as follows: In Section \ref{appatb}, we fix our notation and conventions and recall some key feature of the theoretical framework adopted. In Section \ref{rev}, we first review metric-affine $F(R,\mathcal{D})$ gravity, furnishing the necessary theoretical background and the relevant equations describing the theory. Then, we discuss the particular case $F=R+\alpha \mathcal{D}^2$, assuming linear couplings between the general affine connection and the matter fields (minimal coupling, for self-consistency of the construction) and fixing the torsion vector to zero by means of a Lagrange multiplier in order to break the projective invariance of the theory and allow, on-shell, for a nonvanishing dilation current. 
We study the field equations and derive the (modified) Friedmann equations of the model, in a Friedmann-Lema\^{i}tre-Robertson-Walker (FLRW) background, in the presence of a perfect cosmological hyperfluid (i.e., a classical continuous medium carrying hypermomentum). Section \ref{conclu} is devoted to the conclusions and final remarks.

\section{Theoretical background, notation and conventions}\label{appatb}

We consider a four-dimensional non-Riemannian manifold endowed with a metric and a general affine connection, $\left( \mathcal{M}, g,\nabla \right)$. Our convention for the metric signature is mostly plus: $\left(-,+,+,+\right)$. We use minuscule Greek letters to denote spacetime indices, that is $\mu, \nu,\ldots=0,1,2,3$.
The definition of the covariant derivative of, e.g., a vector $v^\lambda$ is
\begin{equation}
\nabla_\nu v^\lambda = \partial_\nu v^\lambda  + {\Gamma^\lambda}_{\mu \nu} v^\mu \,,
\end{equation}
where ${\Gamma^\lambda}_{\mu \nu}$ is a general affine connection. The latter can be decomposed as
\begin{equation}\label{gendecompaffconn}
{\Gamma^\lambda}_{\mu\nu} = \tilde{\Gamma}^\lambda_{\phantom{\lambda}\mu\nu} + {N^\lambda}_{\mu\nu}\,,
\end{equation}
where
\begin{equation}\label{distortion}
{N^\lambda}_{\mu\nu} = \frac12 g^{\rho\lambda}\left(Q_{\mu\nu\rho} + Q_{\nu\rho\mu}
- Q_{\rho\mu\nu}\right) - g^{\rho\lambda}\left(S_{\rho\mu\nu} +
S_{\rho\nu\mu} - S_{\mu\nu\rho}\right)
\end{equation}
is the distortion tensor,
\begin{equation}\label{lcconn}
\tilde{\Gamma}^\lambda_{\phantom{\lambda}\mu\nu} = \frac12 g^{\rho\lambda}\left(\partial_\mu 
g_{\nu\rho} + \partial_\nu g_{\rho\mu} - \partial_\rho g_{\mu\nu}\right)
\end{equation}
the Levi-Civita connection, and
\begin{equation}
\begin{split}
{S_{\mu\nu}}^\rho & := {\Gamma^\rho}_{[\mu\nu]}\,, \\ 
Q_{\lambda\mu\nu} & := -\nabla_\lambda g_{\mu\nu} = 
-\partial_\lambda g_{\mu\nu} + {\Gamma^\rho}_{\mu\lambda} g_{\rho\nu} +
{\Gamma^\rho}_{\nu\lambda}g_{\mu\rho} 
\end{split}
\end{equation}
the torsion and non-metricity tensors, respectively. In four spacetime dimensions, the trace decomposition of the latter reads
\begin{equation}\label{dectorandnm}
\begin{split}
{S_{\lambda\mu}}^\nu & = \frac{2}{3} \delta_{[\mu}^{\nu} S_{\lambda]} + \frac{1}{6} \varepsilon_{\lambda \mu \kappa \rho} g^{\kappa \nu} t^\rho + {Z_{\lambda\mu}}^\nu \,,  \\
Q_{\lambda\mu\nu} & = \frac{5}{18} Q_\lambda g_{\mu\nu} - \frac19 q_\lambda g_{\mu\nu} +
\frac49 g_{\lambda(\nu}q_{\mu)} - \frac19 g_{\lambda(\nu} Q_{\mu)} + \Omega_{\lambda\mu\nu} \,, 
\end{split}
\end{equation}
where $Q_\lambda := {Q_{\lambda \mu}}^\mu$ and $q_\nu := {Q^\mu}_{\mu\nu}$ are the non-metricity vectors, $S_\lambda :={S_{\lambda \sigma}}^{\sigma}$ is the torsion vector, and $t^\rho := \varepsilon^{\rho \lambda \mu \nu} S_{\lambda \mu \nu}$ is the torsion pseudo-vector. On the other hand, ${Z_{\lambda\mu}}^\nu$ (with $Z_{\lambda \mu \nu} = \frac{4}{3} Z_{[\lambda (\mu]\nu)}$, $\epsilon^{\lambda \mu \nu \rho} Z_{\lambda \mu \nu}=0$) and $\Omega_{\lambda\mu\nu}$ are the traceless parts of torsion and non-metricity, respectively. We denote by $\varepsilon^{\mu \nu \alpha \beta}= \frac{1}{\sqrt{-g}} \epsilon^{\mu \nu \alpha \beta}$ the Levi-Civita tensor, while $\epsilon^{\mu \nu \alpha \beta}$ is the Levi-Civita symbol.
Let us also introduce here the Palatini tensor, which is defined as
\begin{equation}\label{palatinidefin}
{P_{\lambda}}^{\mu\nu} := -\frac{\nabla_{\lambda}(\sqrt{-g}g^{\mu\nu})}{\sqrt{-g}}+\frac{\nabla_{\sigma}(\sqrt{-g}g^{\mu\sigma})\delta^{\nu}_{\lambda}}{\sqrt{-g}} +2(S_{\lambda}g^{\mu\nu}-S^{\mu}\delta_{\lambda}^{\nu}+g^{\mu\sigma}{S_{\sigma\lambda}}^{\nu}) 
\end{equation}
and fulfills
\begin{equation}\label{palatinitraceless}
{P_{\mu}}^{\mu\nu}=0 \,.
\end{equation}
One can then prove that the Palatini tensor can be written in terms of torsion and non-metricity as \cite{Iosifidis:2019dua}
\begin{equation}\label{palatinitornonmet}
\begin{split}
{P_{\lambda}}^{\mu\nu} & = \delta^\nu_\lambda \left(q^\mu - \frac{1}{2} Q^\mu - 2 S^\mu \right) + g^{\mu \nu} \left( \frac{1}{2} Q_\lambda + 2 S_\lambda \right) - \left( {Q_\lambda}^{\mu \nu} + 2 {S_\lambda}^{\mu \nu} \right) \\
& = - {\Omega_\lambda}^{\mu \nu} + \frac{1}{3} g^{\mu \nu} \left( \frac{2}{3} Q_\lambda + \frac{1}{3} q_\lambda + 4 S_\lambda \right) + \frac{1}{9} \delta_\lambda^\nu \left(-4 Q^\mu + 7 q^\mu \right) + \frac{1}{9} \delta_\lambda^\mu  \left( \frac{1}{2} Q^\mu - 2 q^\nu \right) \\
& - \frac{1}{3} {\varepsilon_\lambda}^{\mu \nu \rho} t_\rho - 2 {Z_\lambda}^{\mu \nu} \,.
\end{split}
\end{equation}
Our definition of the Riemann tensor for the general affine connection ${\Gamma^\lambda}_{\mu \nu}$ is
\begin{equation}\label{defRiem}
{R^\mu}_{\nu \alpha \beta} := 2 \partial_{[\alpha} {\Gamma^\mu}_{|\nu|\beta]} + 2 {\Gamma^\mu}_{\rho [\alpha} {\Gamma^\rho}_{|\nu |\beta]} \,.
\end{equation}
Correspondingly, $R_{\mu \nu}:={R^\rho}_{\mu \rho \nu}$ and $R:=g^{\mu \nu} R_{\mu \nu}$ are, respectively, the Ricci tensor and the scalar curvature of $\Gamma$.
The decomposition of the scalar curvature $R$ in terms of the Riemannian scalar curvature $\tilde{R}=g^{\mu \nu}\tilde{R}_{\mu \nu}$ (where $\tilde{R}_{\mu \nu}$ is the Ricci tensor of the Levi-Civita connection $\tilde{\Gamma}$) plus the non-Riemannian contributions is given by
\begin{equation}\label{ricciscaldec}
R = \tilde{R} + T + Q + 2 Q_{\alpha \mu \nu} S^{\alpha \mu \nu} + 2 S_\mu \left( q^\mu - Q^\mu \right) + \tilde{\nabla} \left( q^\mu - Q^\mu - 4 S^\mu \right) \,,
\end{equation}
where $\tilde{\nabla}$ is the Levi-Civita covariant derivative and where we have defined
\begin{equation}\label{tornonmetscals}
\begin{split}
T & := S_{\mu \nu \alpha} S^{\mu \nu \alpha} - 2 S_{\mu \nu \alpha} S^{\alpha \mu \nu} - 4 S_\mu S^\mu \,, \\
Q & := \frac{1}{4} Q_{\alpha \mu \nu} Q^{\alpha \mu \nu} - \frac{1}{2} Q_{\alpha \mu \nu} Q^{\mu \nu \alpha} - \frac{1}{4} Q_\mu Q^\mu + \frac{1}{2} Q_\mu q^\mu \,,
\end{split}
\end{equation}
which are torsion and non-metricity scalars.

\subsection{Non-Riemannian FLRW spacetime}

Let us also recall some relevant expressions obtained in a homogeneous, non-Riemannian FLRW spacetime. Consider a homogeneous, flat FLRW cosmology, where the line element is
\begin{equation}
ds^2 = - dt^2 + a^2(t) \delta_{ij} dx^i dx^j \,,
\end{equation}
with $i,j,\ldots=1,2,3$ and scale factor $a(t)$. The Hubble parameter is
\begin{equation}
H:= \frac{\dot{a}}{a}
\end{equation}
and the projection tensor projecting objects on the space orthogonal to the normalized four-velocity $u^\mu$ (such that $u^\mu=\delta^\mu_0=(1,0,0,0)$ and $u_\mu u^\mu=-1$) is
\begin{equation}\label{projop}
h_{\mu \nu} := g_{\mu \nu} + u_\mu u_\nu = h_{\nu \mu} \,.
\end{equation}
We also define the temporal derivative
\begin{equation}\label{tempder}
\dot{} = u^\alpha \nabla_\alpha \,.
\end{equation}
The general affine connection $\Gamma$ in a non-Riemannian FLRW spacetime in $1+3$ dimensions can be written as \cite{Iosifidis:2020gth}
\begin{equation}\label{connFLRW}
{\Gamma^\lambda}_{\mu \nu} = \tilde{\Gamma}^\lambda_{\phantom{\lambda}\mu \nu} + X(t) u^\lambda h_{\mu \nu} + Y(t) u_\mu {h^\lambda}_\nu + Z(t) u_\nu {h^\lambda}_\mu + V(t) u^\lambda u_\mu u_\nu + {\varepsilon^\lambda}_{\mu \nu \rho} u^\rho W(t) \,,
\end{equation}
where, in particular, the nonvanishing components of the Levi-Civita connection read
\begin{equation}
\tilde{\Gamma}^0_{\phantom{0}ij} = \tilde{\Gamma}^0_{\phantom{0}ji} = \dot{a} a \delta_{ij} \,, \quad \tilde{\Gamma}^i_{\phantom{i}j0} = \tilde{\Gamma}^i_{\phantom{i}0j} = \frac{\dot{a}}{a} \delta^i_j = H \delta^i_j \,. 
\end{equation}
The torsion and non-metricity tensors can be written as
\begin{equation}\label{tornonmetFLRW}
\begin{split}
S_{\mu \nu \alpha} & = 2 u_{[\mu} h_{\nu]\alpha} \Phi(t) + \varepsilon_{\mu \nu \alpha \rho} u^\rho P(t) \,, \\
Q_{\alpha \mu \nu} & = A(t) u_\alpha h_{\mu \nu} + B(t) h_{\alpha(\mu} u_{\nu)} + C(t) u_\alpha u_\mu u_\nu \,,
\end{split}
\end{equation}
respectively. The functions $X(t)$, $Y(t)$, $Z(t)$, $V(t)$, $W(t)$ in \eqref{connFLRW} and $\Phi(t)$, $P(t)$, $A(t)$, $B(t)$, $C(t)$ in \eqref{tornonmetFLRW} describe non-Riemannian cosmological effects and give, together with the scale factor, the cosmic evolution of the non-Riemannian background geometry. 
Recalling the decomposition \eqref{gendecompaffconn}, one can prove that
\begin{equation}
2(X+Y) = B \,, \quad 2 Z = A \,, \quad 2 V = C \,, \quad 2 \Phi = Y - Z \,, \quad P=W \,,
\end{equation}
which may also be inverted, yielding
\begin{equation}
W = P \,, \quad V= \frac{C}{2} \,, \quad Z = \frac{A}{2} \,, \quad Y = 2 \Phi + \frac{A}{2} \,, \quad X = \frac{B}{2} - 2 \Phi - \frac{A}{2} \,.
\end{equation}
Besides, the torsion and non-metricity scalars defined in \eqref{tornonmetscals} now become, respectively,
\begin{equation}\label{tornmscalFLRM}
\begin{split}
T & = 24 \Phi^2 - 6 P^2 \,, \\
Q & = \frac{3}{4} \left[ 2 A^2 + B (C-A) \right] \,.
\end{split}
\end{equation}
Finally, using the post-Riemannian decomposition of the curvature scalar and the above expressions of the torsion and non-metricity scalars, we find
\begin{equation}\label{prexpcosm}
R = \tilde{R} + 6 \left[ \frac{1}{4} A^2 + 4 \Phi^2 + \Phi (2A-B) \right] + \frac{3}{4} B (C-A) - 6 P^2 + \frac{3}{\sqrt{-g}} \partial_\mu \left[ \sqrt{-g} u^\mu \left( \frac{B}{2} - A - 4 \Phi \right) \right] \,,
\end{equation}
where
\begin{equation}
\tilde{R} = 6 \left[ \frac{\ddot{a}}{a} + \left( \frac{\dot{a}}{a} \right)^2 \right]
\end{equation}
is the usual Riemannian part, namely the Ricci scalar of the Levi-Civita connection $\tilde{\Gamma}$ expressed in terms of the scale factor and its derivatives.

\section{Metric-affine $F(R,\mathcal{D})$ gravity}\label{rev}

Let us now briefly review metric-affine $F(R,\mathcal{D})$ gravity.
We start by recalling that, as far as the matter content in MAG is concerned, besides the usual (\textit{metrical}) energy-momentum
tensor, which is defined as
\begin{equation}
T_{\mu \nu} := - \frac{2}{\sqrt{-g}} \frac{\delta S_{\text{m}}}{\delta g^{\mu \nu}} = - \frac{2}{\sqrt{-g}} \frac{\delta \left( \sqrt{-g} \mathcal{L}_{\text{m}} \right)}{\delta g^{\mu \nu}} \,,
\end{equation}
where $\mathcal{L}_{\text{m}}$ is the matter Lagrangian, we also have a nontrivial dependence of $\mathcal{L}_{\text{m}}$ on the general affine connection ${\Gamma^\lambda}_{\mu \nu}$. The variation of the matter part of the action with respect to ${\Gamma^\lambda}_{\mu \nu}$ defines the hypermomentum tensor,
\begin{equation}
{\Delta_\lambda}^{\mu \nu} :=  - \frac{2}{\sqrt{-g}} \frac{\delta S_{\text{m}}}{\delta {\Gamma^\lambda}_{\mu \nu}} = - \frac{2}{\sqrt{-g}} \frac{\delta \left( \sqrt{-g} \mathcal{L}_{\text{m}}\right)}{\delta {\Gamma^\lambda}_{\mu \nu}} \,.
\end{equation}
The hypermomentum has a direct physical interpretation when split into its irreducible pieces of spin, dilation, and shear \cite{Hehl:1976kt,Hehl:1976kt2,Hehl:1976kv}. In particular, in \cite{Iosifidis:2021kqo,Myrzakulov:2021vel} it was considered the metric-affine action
\begin{equation}\label{act1}
    \mathcal{S}_{F(R,\mathcal{D})} = \frac{1}{2 \kappa} \int \sqrt{-g} d^4 x \left[ F(R,\mathcal{D}) + 2 \kappa \mathcal{L}_{\text{m}} \right] \,,
\end{equation}
where $\kappa = 8 \pi G$ is the gravitational constant, and $F(R,\mathcal{D})$ is a general function of the curvature scalar $R$ of the general affine connection ${\Gamma^\lambda}_{\mu \nu}$ involving torsion and non-metricity and $\mathcal{D}$ is a hypermomentum contribution that was first introduced in \cite{Iosifidis:2021kqo}. This quantity is defined as
\begin{equation}
    \mathcal{D} := \frac{1}{\sqrt{-g}} \partial_\nu \left( \sqrt{-g} \Delta^\nu \right) = \tilde{\nabla}_\nu \Delta^\nu \,,
\end{equation}
where 
\begin{equation}
\Delta^\nu := {\Delta_\mu}^{\mu \nu} 
\end{equation}
is the dilation current. Thus, in this generalized metric-affine setup one can consider also the presence of the hypermomentum
analog of the (metrical) energy-momentum trace. Indeed, let us notice that the divergence of the dilation current is analogous to the trace $\mathcal{T}:=g^{\mu \nu}T_{\mu \nu}$ of the energy-momentum tensor: They both appear in the expression of the trace $t:=g^{\mu \nu}t_{\mu \nu}$ of the so-called \textit{canonical} energy-momentum tensor, that is\footnote{The canonical energy-momentum tensor, in general, is not symmetric and is given by the variation of the gravitational action with respect to the vielbein, as one can work in the equivalent formalism based on the vielbein ${e_{\mu}}^c$ and spin connection $\omega_{\mu | a b}$, where $a,b,\ldots$ are Lorentz (i.e., tangent) indices (see, e.g., \cite{Iosifidis:2020gth}). We have the usual relation $g_{\mu \nu}= {e_\mu}^a {e_\nu}^b \eta_{ab}$ connecting metric and vielbein, being $\eta_{ab}$ the tangent space flat Minkowski metric. The identity $$\nabla_\nu {e_\mu}^a = 0 = \partial_\nu {e_\mu}^a - {\Gamma^\rho}_{\mu \nu} {e_\rho}^a + \omega^{\phantom{\nu} a}_{\nu \phantom{a} b} {e_\mu}^b$$ connects the two formalisms.}
\begin{equation}\label{cemt}
{t^\mu}_c := \frac{1}{\sqrt{-g}} \frac{\delta S_{\text{m}}}{\delta {e_\mu}^c} = \frac{1}{\sqrt{-g}} \frac{\delta \left( \sqrt{-g} \mathcal{L}_{\text{m}} \right)}{\delta {e_\mu}^c} \,,
\end{equation}
as
\begin{equation}\label{tracetTandD}
    t = \mathcal{T} + \frac{1}{2\sqrt{-g}} \partial_\nu \left(\sqrt{-g} \Delta^\nu \right) \,,
\end{equation}
or, in terms of the Levi-Civita covariant derivative,
\begin{equation}
    t = \mathcal{T} + \frac{1}{2} \tilde{\nabla}_\nu \Delta^\nu \,.
\end{equation}
In this sense, the energy-momentum trace $\mathcal{T}$ and the divergence of $\Delta^\nu$ can be placed on an equal footing, and the scalar $\mathcal{D}$ obtained by the divergence of the dilation current stands as the analog of the trace of the energy-momentum tensor, but for the hypermomentum tensor. It should be stressed out  that, in general, $\mathcal{D}$ depends both on the metric and on the connection.

Let us now recall the field equations of the family of theories given by the the action \eqref{act1}. The variation of the action with respect to the metric and the general affine connection gives the following set of field equations, respectively:\footnote{We adopt the notation $F'_X := \frac{dF}{dX}$ to denote the derivative of $F$ with respect to any scalar $X$ of which $F$ is function.}
\begin{equation}
\begin{split}
& - \frac{1}{2} g_{\mu \nu} F + F'_R R_{(\mu \nu)} + F'_{\mathcal{D}} M_{\mu \nu} = \kappa T_{\mu \nu} \,, \\
& {P_\lambda}^{\mu \nu} (F'_R) - {M_\lambda}^{\mu \nu \rho} \partial_\rho F'_{\mathcal{D}} = \kappa {\Delta_\lambda}^{\mu \nu} \,,
\end{split}
\end{equation}
where ${P_\lambda}^{\mu \nu} (F'_R)$ is the modified Palatini tensor,
\begin{equation}\label{modpala}
{P_\lambda}^{\mu \nu} (F'_R) := - \frac{\nabla_\lambda \left(\sqrt{-g} F'_R g^{\mu \nu} \right)}{\sqrt{-g}} + \frac{\nabla_\alpha \left( \sqrt{-g} F'_R g^{\mu \alpha}\delta^\nu_\lambda \right)}{\sqrt{-g}} + 2 F'_R \left( S_\lambda g^{\mu \nu} - S^\mu \delta^\nu_\lambda - {S_\lambda}^{\mu \nu} \right) \,,
\end{equation}
with $S_\lambda :={S_{\lambda \sigma}}^{\sigma}$ the torsion vector, and
\begin{align}
    & M_{\mu \nu} := \frac{\delta \mathcal{D}}{\delta g^{\mu \nu}} = M_{\nu \mu} \,, \label{m2tens} \\
    & {M_\lambda}^{\mu \nu \rho} := \frac{\delta \Delta^\rho}{\delta {\Gamma^\lambda}_{\mu \nu}} \,. \label{m4tens}
\end{align}
Note that in the simpler case in which matter does not couple to the connection (that is, for instance, in the case of a classical perfect fluid with no inner structure) we have ${\Delta_\lambda}^{\mu \nu}=0$ and, consequently, no hypermomentum contribution in the action and field equations. The form of the tensor ${M_\lambda}^{\mu \nu \rho}$ depends crucially on how we couple matter to gravity. More specifically, if there are only linear couplings between the connection and the matter fields, the aforementioned tensor identically vanishes, while for quadratic couplings it is linear in the connection. For self-consistency of the derivation, and in particular of eq. \eqref{act1}, in this work we consider only minimal coupling with the affine connection (that is, linear couplings, such that higher-order matter-affine connection interactions are absent).

In \cite{Iosifidis:2021kqo} the cosmology of (an extension of) this model was analyzed for a function $F$ that is linear, in particular, in $R$ and $\mathcal{D}$. However, since $\sqrt{-g}\mathcal{D}$ is a total divergence, the dilation current does not contribute to the field equations when included linearly, and can be trivially neglected. We shall now discuss a more involved case, that is metric-affine $F(R,\mathcal{D})=R+\alpha \mathcal{D}^2$ gravity, where the function $F(R,\mathcal{D})$ is still linear in $R$, but quadratic in $\mathcal{D}$.

\subsection{Dilation current in metric-affine $F=R+\alpha \mathcal{D}^2$ gravity}\label{ourmodel}

Let us proceed by focusing on the case in which $F(R,\mathcal{D})=R+\alpha \mathcal{D}^2$, where $\alpha$ is a free parameter, in the theory \eqref{act1},\footnote{In fact, one may also start by writing $F = \alpha_0 R + \alpha \mathcal{D}^2$ and then fix the normalization of the theory choosing $\alpha_0 = 1$, ending up with the theory we are considering here.} giving some results also at the cosmological level. 
In the following, we will assume only linear couplings between the connection and the matter fields. However, considering the variation of the total action with respect to the general affine connection, for metric-affine $F(R,\mathcal{D})=R+\alpha \mathcal{D}^2$ gravity this would lead, on-shell, to a vanishing dilation current (one can prove this by taking into account the fact that the Palatini tensor \eqref{palatinidefin} fulfills ${P_{\mu}}^{\mu \nu}=0$), which cannot be true for any form of matter. This issue appears because of the projective invariance of the scalar curvature $R$ (see also \cite{Iosifidis:2019dua}). To obtain a self-consistent theory with $\Delta^\nu \neq 0$ one needs to break this projective invariance. This can be done, for instance, by adding extra terms in the action that do not respect the projective invariance. In particular, driven by the fact that in \cite{Hehl:1976my} the dilation current was associated to the non-metricity vector $Q_\mu := {Q_{\mu \nu}}^\nu$ (sometimes also referred to as the Weyl vector), we add to the theory a Lagrange multiplier that fixes the torsion vector $S_\mu$ to zero. 

Thus, we consider the following metric-affine theory:
\begin{equation}\label{act2}
    \mathcal{S} = \frac{1}{2 \kappa} \int \sqrt{-g} d^4 x \left[ R + \alpha \mathcal{D}^2 + 2 \kappa \left( B_\mu S^\mu + \mathcal{L}_{\text{m}} \right) \right] \,,
\end{equation}
where $B_\mu$ is a Lagrange multiplier. Varying the action with respect to the latter, we get
\begin{equation}\label{zerotorvec}
    S_\mu = 0 \,.
\end{equation}
As now we have $F'_R=1$ and $F'_\mathcal{D}=2 \alpha \mathcal{D}$, using also \eqref{zerotorvec}, the metric field equations becomes
\begin{equation}\label{mefe}
    - \frac{1}{2} g_{\mu \nu} \left(R+\alpha \mathcal{D}^2 \right) + R_{(\mu \nu)} + 2 \alpha \mathcal{D} M_{\mu \nu} = \kappa T_{\mu \nu} \,, 
\end{equation}
where we recall that $M_{\mu \nu}$ is defined in \eqref{m2tens}. Taking the trace of the above equation we get
\begin{equation}\label{mefetrace}
    R = 2 \alpha \mathcal{D} (M-\mathcal{D}) - \kappa \mathcal{T} \,,
\end{equation}
where $M:=g^{\mu \nu}M_{\mu \nu}$. Using the trace equation into \eqref{mefe}, we are left with
\begin{equation}\label{mefetraceless}
    R_{(\mu \nu)} - \frac{1}{4} g_{\mu \nu} R = - 2 \alpha \mathcal{D} \left( M_{\mu \nu} - \frac{1}{4} g_{\mu \nu} M \right) + \kappa \left( T_{\mu \nu} - \frac{1}{4} g_{\mu \nu} \mathcal{T} \right) \,.
\end{equation}
Notice that this equation may also be rewritten as
\begin{equation}
    \mathring{R}_{(\mu \nu)} = - 2 \alpha \mathcal{D} \mathring{M}_{\mu \nu} + \kappa \mathring{T}_{\mu \nu} \,,
\end{equation}
where we have used the symbol $\mathring{}$ to denote traceless tensors.
On the other hand, the connection field equations of the theory boil down to
\begin{equation}\label{confe}
    {P_\lambda}^{\mu \nu} = \kappa \left( {\Delta_\lambda}^{\mu \nu} - B^{[\mu} \delta^{\nu]}_\lambda \right)\,,
\end{equation}
where ${P_\lambda}^{\mu \nu}$ is the Palatini tensor defined in \eqref{palatinidefin} (recall that now $S_\mu=0$).
Taking the $\lambda,\mu$ trace of \eqref{confe} we get (let us recall that the Palatini tensor ${P_\lambda}^{\mu \nu}$ fulfills ${P_\mu}^{\mu \nu} =0$)
\begin{equation}\label{dilationeq}
    B^\nu = - \frac{2}{3} \Delta^\nu \,,
\end{equation}
which express the Lagrange multiplier completely in terms of the dilation current. Note that this means that, in principle, we have that, on-shell, $\Delta^\nu \neq 0$, contrary to the simpler case of metric-affine $F(R)$ gravity (see, e.g., \cite{Iosifidis:2019dua}). This would not have been the case if we had not broken the projective invariance of the theory.
On the other hand, using the above and exploiting the fact that the Palatini tensor can be written in terms of torsion and non-metricity as given in \eqref{palatinitornonmet}, taking the $\lambda,\nu$ and $\mu, \nu$ traces of \eqref{confe}, together with its totally antisymmetric part, after some manipulation we find
\begin{align}
        & Q_\mu = - \kappa \left( \frac{1}{3} \Delta_\mu + \frac{1}{3} \Delta^{(1)}_\mu - \Delta^{(2)}_\mu \right) \,, \\
        & q_\mu = \frac{\kappa}{2} \left( \frac{1}{3} \Delta_\mu + \frac{1}{3} \Delta^{(1)}_\mu + \Delta^{(2)}_\mu \right) \,, \\
        & t_\mu = - \frac{\kappa}{2} \varepsilon_{\mu \nu \rho \sigma} \Delta^{\nu \rho \sigma} \,, \\
        & \Omega_{\lambda \mu \nu} = \frac{\kappa}{18} \left( - 7 g_{\mu \nu} \Delta_\lambda + 10 g_{\lambda (\mu} \Delta_{\nu)} + 5 g_{\mu \nu} \Delta^{(1)}_\lambda - 2 g_{\lambda (\mu} \Delta^{(1)}_{\nu)} + 5 g_{\mu \nu} \Delta^{(2)}_\lambda - 2 g_{\lambda (\mu} \Delta^{(2)}_{\nu)} \right) \nonumber \\
        & \phantom{\Omega_{\lambda \mu \nu}} + \kappa \left( \Delta_{(\mu \nu) \lambda} - \Delta_{\lambda (\mu \nu)} - \Delta_{(\mu | \lambda |\nu)} \right) \,, \\
        & {Z_{\lambda \mu}}^\nu = \frac{\kappa}{3} \left( - \delta_{[\lambda}^\nu \Delta_{\mu]} + \delta_{[\lambda}^\nu \Delta^{(1)}_{\mu]} - {\Delta_{[\lambda \mu]}}^\nu + \Delta_{[\lambda \phantom{\nu} \mu]}^{\phantom{[\lambda} \nu} + 2 {\Delta^\nu}_{[\lambda \mu]} \right) \,,
    \end{align}
where we have defined $\Delta^{(1)}_\mu :={\Delta^\lambda}_{\mu \lambda}$ and $\Delta^{(2)}_\mu :=\Delta^{\phantom{\mu} \lambda}_{\mu \phantom{\lambda} \lambda}$.
Hence, all the non-Riemannian components of the general affine connection ${\Gamma^{\lambda}}_{\mu \nu}$, whose decomposition is reported in Section \ref{appatb}, result to be given in terms of hypermomentum quantities, while we recall that the torsion trace vanishes in our setup ($S_\mu=0$). Notice that the divergence of the dilation current contributes only to the metric field equations of the theory.

\subsubsection{Cosmological aspects of the model: Modified Friedmann equations}\label{cosman}

In this section we study the cosmology of metric-affine $F=R+\alpha \mathcal{D}^2$ gravity, considering a homogeneous FLRW background in the presence of torsion and non-metricity (see Section \ref{appatb}), deriving the Friedmann equations of the model.

Taking the trace of the metric field equations \eqref{mefe} (that is, considering \eqref{mefetrace}), using \eqref{tornonmetFLRW} and the post-Riemannian cosmological expansion \eqref{prexpcosm}, after some cumbersome calculations we finally arrive at
\begin{equation}\label{firstF}
    \frac{\ddot{a}}{a} + \left(\frac{\dot{a}}{a} \right)^2 + 4 \Phi^2 - P^2 + \frac{1}{8} \left[ 2 A^2 + B \left( C - A \right) \right] + \Phi \left(2A-B\right) + \dot{f} + 3 H f = \frac{\alpha}{3} \mathcal{D} \left( M - \mathcal{D} \right) - \frac{\kappa}{6} \mathcal{T} \,,
\end{equation}
with
\begin{equation}\label{ffunc}
    f := \frac{1}{2} \left(\frac{B}{2} - A - 4 \Phi \right) \,,
\end{equation}
where $a=a(t)$ is the scale factor of the Universe and $H:=\frac{\dot{a}}{a}$ is Hubble parameter, while $\Phi=\Phi(t)$, $P=P(t)$, $A=A(t)$, $B=A(t)$, and $C=C(t)$ are functions appearing in the FLRW expressions of torsion and non-metricity given in \eqref{tornonmetFLRW}.
Equation \eqref{firstF} is a modification of the first Friedmann equation in our cosmological and geometric setup. Note, in particular, that $\mathcal{D}$ contributes to the right-hand side of this equation. 

On the other hand, the general form of the second Friedmann equation (also known as acceleration equation) in a non-Riemannian background was derived in \cite{Iosifidis:2020zzp} and reads
\begin{equation}\label{secondF}
\frac{\ddot{a}}{a} = - \frac{1}{3} R_{\mu \nu} u^\mu u^\nu + 2 \left( \frac{\dot{a}}{a} \right) \Phi + 2 \dot{\Phi} + \left( \frac{\dot{a}}{a} \right) \left( A + \frac{C}{2} \right) + \frac{\dot{A}}{2} - \frac{A^2}{4} - \frac{1}{4} A C - A \Phi - C \Phi \,,
\end{equation}
where $u^\mu$ is the normalized four-velocity.
One could then proceed by contracting \eqref{mefe} with $u^\mu u^\nu$ in order to eliminate the first term appearing in the right-hand side of \eqref{secondF} (namely, $R_{\mu \nu} u^\mu u^\nu$) and express everything in terms of the scale factor and the torsion and non-metricity variables. However, in order to better analyze in depth this cosmological model one should consider an
appropriate form of matter for which both the (metrical) energy-momentum and hypermomentum tensors respect the cosmological principle. In \cite{Iosifidis:2020gth} it was introduced a cosmological fluid that fulfills this requirement and was named perfect cosmological hyperfluid. In particular, the hypermomentum part of such fluid sources the torsion and non-metricity variables by means of the connection field equations. Besides, scalar fields coupled to the general affine connection can be regarded as sub-cases of the aforementioned fluid description.
Hence, we are now going to discuss metric-affine $F=R+\alpha \mathcal{D}^2$ gravity in the presence of a perfect cosmological hyperfluid, demanding also homogeneity of the cosmological setting.

A hyperfluid is described in terms of the metrical energy-momentum tensor
\begin{equation}\label{metrenmomform}
T_{\mu \nu} = \rho u_\mu u_\nu + p h_{\mu \nu} 
\end{equation}
along with the (symmetric) canonical energy-momentum tensor
\begin{equation}\label{canonenmomform}
t_{\mu \nu} = \rho_c u_\mu u_\nu + p_c h_{\mu \nu} \,,
\end{equation}
where $\rho$ and $p$ are the usual density and pressure of the perfect fluid component, while $\rho_c$ and $p_c$ are the canonical (net) density and canonical pressure of the hyperfluid, respectively. 
Besides, in four spacetime dimensions the perfect hyperfluid is characterized by a hypermomentum tensor of the form \cite{Iosifidis:2020gth} 
\begin{equation}\label{hypermomform}
\Delta_{\alpha \mu \nu} = \phi(t) h_{\mu \alpha} u_\nu + \chi(t) h_{\nu \alpha} u_{\mu} + \psi(t) u_{\alpha} h_{\mu \nu} + \omega(t) u_\alpha u_\mu u_\nu + \varepsilon_{\alpha \mu \nu \rho} u^\rho \zeta(t) \,,
\end{equation}
where the functions $\phi(t)$, $\chi(t)$, $\psi(t)$, $\omega(t)$, and $\zeta(t)$ characterize the microscopic properties of the hyperfluid which, upon use of the connection field equations, act as sources of the non-Riemannian background. The tensors \eqref{metrenmomform}, \eqref{canonenmomform}, and \eqref{hypermomform} respect spatial isotropy and are subject, in general, to conservation laws. 
If the fluid considered is of the so-called ``hypermomentum preserving'' type, that is if the metrical energy-momentum tensor coincides with the canonical one, $t^{\mu \nu} = T^{\mu \nu}$, then, besides having $\rho_c = \rho$ and $p_c = p$, we also get $t=\mathcal{T}$, namely, in this case, the trace of the canonical and metrical energy-momentum tensors coincide. Looking at \eqref{tracetTandD} we can see that this implies $\mathcal{D}=0$. Such a feature, in the theory we are now considering, would remove the $\mathcal{D}$ contribution appearing into the metric field equations.\footnote{However, in more general metric-affine $F(R,\mathcal{D})$ theories such trivialization may not occur in the hypermomentum preserving case.} Thus, let us consider $t^{\mu \nu} \neq T^{\mu \nu}$, that is $\rho_c \neq \rho$ and $p_c \neq p$.

Since we are mainly interested in the modification to the cosmological equations sourced by dilation, we will focus on the case in which only this specific part of the hypermomentum tensor is switched on, that is pure dilation hypermomentum. In four spacetime dimensions the hypermomentum tensor can be decomposed as \cite{Iosifidis:2020upr}
\begin{equation}
\Delta_{\alpha \mu \nu} = \Delta_{[\alpha \mu] \nu} + \frac{1}{4} g_{\alpha \mu} \Delta_\nu + \breve{\Delta}_{\alpha \mu \nu} \,,
\end{equation}
where the first term on the right-hand side represents the spin part, $\Delta_\nu$, as we have already said, is the dilation, and $\breve{\Delta}_{\alpha \mu \nu}$ the shear (traceless symmetric part of the hypermomentum tensor). Then, given the most general form \eqref{hypermomform} of hypermomentum compatible with the cosmological principle (i.e. respecting both isotropy and homogeneity), the spin, dilation and shear parts read
\begin{align}
& \Delta_{[\alpha \mu ] \nu} = \left( \psi - \chi \right) u_{[\alpha} h_{\mu]\nu} + \epsilon_{\alpha \mu \nu \rho} u^\rho \zeta \,, \label{hspin} \\
& \Delta_\nu := \Delta_{\alpha \mu \nu} g^{\alpha \mu} = \left( 3 \phi - \omega \right) u_\nu \,, \label{hdilation} \\
& \breve{\Delta}_{\alpha \mu \nu} = \Delta_{(\alpha \mu )\nu} - \frac{1}{4} g_{\alpha \mu} \Delta_\nu = \frac{\left( \phi + \omega \right)}{4} \left( h_{\alpha \mu} + 3 u_\alpha u_\mu \right) u_\nu + \left( \psi + \chi \right) u_{(\mu} h_{\alpha ) \nu} \,, \label{hshear}
\end{align}
respectively. Thus, in the case of pure dilation hypermomentum we are left with
\begin{equation}
    \psi = 0 \,, \quad \chi = 0 \,, \quad \zeta = 0 \,, \quad \omega = - \phi \,,
\end{equation}
that is
\begin{equation}
\begin{split}
    & \Delta_\nu = 4 \phi u_\nu \,, \\
    & \Delta_{\alpha \mu \nu} = \frac{1}{4} g_{\alpha \mu} \Delta_\nu = \phi g_{\alpha \mu} u_\nu = \phi \left(h_{\alpha \mu} u_\nu - u_\alpha u_\mu u_\nu \right) \,. \label{puredilhypcosm}
\end{split}    
\end{equation}
Furthermore, we will assume the following barotropic equations of state to hold:
\begin{equation}\label{barotropiceos}
p= w \rho \,, \quad p_c = w_c \rho_c \,,
\end{equation}
where $w$ and $w_c$ are the associated barotropic index.

Before proceeding, let us also comment on the tensor $M_{\mu \nu}$ appearing into the metric field equations. The most general cosmological expression for it is
\begin{equation}\label{Mmunuansatz}
    M_{\mu \nu} = \theta u_\mu u_\nu + \sigma h_{\mu \nu} \,, 
\end{equation}
where $\theta=\theta(t)$ and $\sigma=\sigma(t)$ are functions characterizing $M_{\mu \nu}$. Their form can be found using the definition of $M_{\mu\nu}$ and the fact we are in a non-Riemannian FLRW spacetime. We obtain
\begin{equation}
M_{00}=\frac{\delta \mathcal{D}}{\delta g^{00}} \Big \vert_{N=1} =-\frac{1}{2}\mathcal{D} \,,
\end{equation}
where $N$ denotes the lapse function, and
\begin{equation}
M_{ij}=\frac{\delta \mathcal{D}}{\delta g^{ij}}=2 \left( \dot{\phi} + 3 H \phi \right) g_{ij} \,,
\end{equation}
implying $\theta=-\mathcal{D}/2$ and $\sigma=2 \left( \dot{\phi} + 3 H \phi \right)=-\theta$. Note that, in the present cosmological setup, we have
\begin{equation}\label{Dilcosmexpr}
    \mathcal{D} = 4 \left( \dot{\phi} + 3 H \phi \right) \,.
\end{equation}
Therefore, we are left with
\begin{equation}\label{Mnunuexplform}
M_{\mu\nu}=-\frac{\mathcal{D}}{2}u_{\mu}u_{\nu}+2 \left( \dot{\phi} + 3 H \phi \right)h_{\mu\nu} = \frac{\mathcal{D}}{2} \left( h_{\mu \nu} - u_{\mu}u_{\nu} \right) \,. 
\end{equation}
In the following we will take this to be the cosmological expression of $M_{\mu\nu}$.

Within the above cosmological setup, using the cosmological expressions for torsion and non-metricity given in Section \ref{appatb}, from the field equation of the Lagrange multiplier we find
\begin{equation}
    \Phi = 0 \,,
\end{equation}
while from the connection field equations, taking all the possible contractions, we get
\begin{equation}
    B^\nu = - \frac{8}{3} \phi u^\nu \,,
\end{equation}
namely the Lagrange multiplier is given in terms of the hypermomentum source $\phi$, together with
\begin{align}
    & P = 0 \,, \\
    & A = C = - \frac{\kappa}{3} \phi \,, \\
    & B = \frac{2 \kappa}{3} \phi \,.
\end{align}
Thus, at cosmological level, in the presence of a perfect hyperfluid with pure dilation hypermomentum, we are left with a vanishing torsion ($S_{\mu \nu \alpha}=0$, as actually expected at this point), while the non-metricity is completely given in terms of the hypermomentum source $\phi$,
\begin{equation}
    Q_{\alpha \mu \nu} = - \frac{\kappa}{3} \phi \left( u_\alpha h_{\mu \nu} + u_\alpha u_\mu u_\nu \right) + \frac{2 \kappa}{3} \phi h_{\alpha (\mu} u_{\nu)} \,,
\end{equation}
which implies
\begin{equation}
    Q_\mu = - \frac{2 \kappa}{3} \phi u_\mu = - \frac{1}{2} q_\mu \,, \quad \Omega_{\alpha \mu \nu} = 0 \,.
\end{equation}
We shall now study the Friedmann equations \eqref{firstF} and \eqref{secondF}.

Taking into account all of the above, from the metric field equations \eqref{mefe}, upon contraction with $u^\mu u^\nu$ and use of \eqref{mefetrace}, we find
\begin{equation}
    R_{\mu \nu} u^\mu u^\nu = \alpha \left[\frac{1}{2} \mathcal{D}^2 - \mathcal{D} \left(\theta + 3 \sigma \right) \right] + \frac{\kappa}{2} \left( 1 + 3 w \right) \rho \,.
\end{equation}
Then, using \eqref{Dilcosmexpr}, the expression for $R_{\mu \nu} u^\mu u^\nu$ becomes
\begin{equation}
    R_{\mu \nu} u^\mu u^\nu = 4 \alpha \left( \dot{\phi} + 3 H \phi \right) \left[ 2 \left( \dot{\phi} + 3 H \phi \right) - \left(\theta + 3 \sigma \right) \right] + \frac{\kappa}{2} \left( 1 + 3 w \right) \rho \,.
\end{equation}
Plugging this and the results previously obtained back into the acceleration equation, the latter boils down to
\begin{equation}\label{secondFnew}
\begin{split}
    \frac{\ddot{a}}{a} & = - \frac{1}{18} \kappa^2 \phi^2 - \frac{\kappa}{2} \left[ \frac{1}{3} \dot{\phi} + H \phi + \frac{1}{3} \left(1+3w\right) \rho \right] \\
    & + 4 \alpha \left( - \frac{2}{3} \dot{\phi}^2 - 4 \dot{\phi} H \phi - 6 H^2 \phi^2 + \dot{\phi}\sigma + 3 H \phi \sigma + \frac{1}{3} \dot{\phi} \theta + H \phi \sigma \right) \,.
\end{split}    
\end{equation}
Finally, using the expression of the functions $\theta$ and $\sigma$ previously derived, together with \eqref{Dilcosmexpr}, we obtain
\begin{equation}
    \frac{\ddot{a}}{a} = - \frac{\kappa}{6} \rho (1+3w) - \frac{\kappa^2}{18} \phi^2 - \frac{\kappa}{24} \mathcal{D} + \frac{\alpha}{6} \mathcal{D}^2 \,. \label{ddafinal}
\end{equation}
We can then substitute the obtained result for $\frac{\ddot{a}}{a}$ into \eqref{firstF}, in such a way to arrive at the final form of the first (modified) Friedmann equation for the model at hand, which, under the above assumptions, reads
\begin{equation}
    H^2 \left( 1 + 24 \alpha \phi^2 \right) = \frac{\kappa^2}{36} \phi^2 + \kappa \left( - \frac{1}{6} \dot{\phi} - \frac{1}{2} H \phi + \frac{1}{3} \rho \right) + \alpha \left( - \frac{8}{3} \dot{\phi}^2 - 16 \dot{\phi} H \phi - \frac{8}{3} \dot{\phi} \theta - 8 H \phi \theta \right) \,,
\end{equation}
that is, using the cosmological form of $\theta$ and $\sigma$ and \eqref{Dilcosmexpr},
\begin{equation} 
    H^2 = \frac{\kappa}{3} \rho + \frac{\kappa^2}{36} \phi^2 - \frac{\kappa}{24} \mathcal{D} + \frac{\alpha}{6} \mathcal{D}^2 \,. \label{HHsqfinal} 
\end{equation}
Note that nontrivial modification to the cosmology of the theory induced by the presence of $\mathcal{D}$ may appear only in the case in which $\rho \neq \rho_c$. Furthermore, we observe that, in any possible solution of the above equation (which should also involve the explicit expression of the hyperfluid conservation laws), the effect of $\phi$ is to enhance the total energy density, while the term with $\mathcal{D}^2$ may in fact enhance or diminish the total energy density depending on the sign of the parameter $\alpha$ characterizing the model $F(R,\mathcal{D})=R+\alpha \mathcal{D}^2$.

\section{Conclusions}\label{conclu}

Modified theories of gravity can be discussed in the MAG framework, where coupling with matter is particularly interesting to be studied, as the matter Lagrangian is considered to depend on the connection as well. In this work, we have considered a particular family of MAG theories, that is $F(R,\mathcal{D})$ metric-affine gravity, where $\mathcal{D}$ is the divergence of the dilation current. The latter appears in the hypermomentum tensor and relates the trace of the so-called canonical and metrical energy-momentum tensors. It is in this sense that $F(R,\mathcal{D})$ gravity may be said to be the hypermomentum analog of $F(R,\mathcal{T})$ gravity. 
In the case in which the function $F(R,\mathcal{D})$ is linear in $\mathcal{D}$, the dilation current does not contribute to the field equations ($\sqrt{-g} \mathcal{D}$ in the action is a total derivative), while in more complicated cases, as we have discussed in the present paper, it can have nontrivial effects, in particular affecting the cosmology of the theory.

More specifically, after reviewing in general but concise terms metric-affine $F(R,\mathcal{D})$ gravity, we have discussed the case $F=R+\alpha \mathcal{D}^2$, assuming linear couplings between the general affine connection and the matter fields (minimal coupling) and consistently breaking the projective invariance of the theory (setting the torsion vector $S_\mu$ to zero by means of a Lagrange multiplier). We have derived the field equations of the theory, showing that the dilation current contributes only to the metric field equations of the model. On the other hand, the general affine connection exhibits non-Riemannian components given in terms of hypermomentum quantities. Only the torsion trace vanishes in our setup (so that projective invariance is broken), while the other non-Riemannian components of the connection are, a priori, non-vanishing.

Subsequently, to study the cosmology of the theory, we have considered a homogeneous FLRW background in the presence of torsion and non-metricity. We have introduced a perfect cosmological hyperfluid, whose hypermomentum part sources torsion and non-metricity by means of the connection field equations. Hence, we have derived the modified Friedmann equations of the $R+\alpha \mathcal{D}^2$ model in this setup, showing that, in particular, the presence of the dilation current $\mathcal{D}$ affects the first modified Friedmann equation.
In performing this analysis, we have also presented the most general cosmological expression for the tensor $M_{\mu \nu}$ appearing into the metric field equations of the model.
We have focused on the case of pure dilation hypermomentum, which is given in terms of a single function $\phi=\phi(t)$ characterizing the microscopic properties of the hyperfluid. This variable, upon use of the connection field equations, acts as source of the non-Riemannian objects. Specifically, we have shown that, at the cosmological level, in the presence of a perfect hyperfluid with pure dilation hypermomentum, the torsion vanishes, while the non-metricity results to be completely determined by the hypermomentum source $\phi$. 
We have found that nontrivial modifications to the cosmology of the theory induced by $\mathcal{D}$ can appear only in the case in which $\rho \neq \rho_c$, that is, only if the energy density of the perfect fluid component is different from the canonical (net) density of the hyperfluid. The cosmological effect of $\phi$ is to always enhance the total density, while the $\mathcal{D}^2$ contribution may increase or decrease the total energy density depending on the sign of the parameter $\alpha$.
It is certainly worth looking for exact solutions involving the effects of $\mathcal{D}$, including also the explicit derivation of the hyperfluid conservation laws.

\section*{Acknowledgments}

We thank D. Iosifidis for guidance and highlighting discussions and suggestions, especially in the first stages of this work.
L.R. would like to thank the DISAT of the Polytechnic of Turin and the INFN for financial support. 
This work was supported by the Ministry of Science and Higher Education of the Republic of Kazakhstan, Grant AP14870191.


\end{document}